\documentclass[11pt]{article}
\usepackage{moriond}
\usepackage[centertags]{amsmath}
\usepackage{amssymb,amsfonts,theorem,stmaryrd}
\usepackage{slashbox}
\usepackage{graphicx}
\usepackage{slashed}
\usepackage[all]{xy}

\newcommand{\M}{\mathcal{M}}
\newcommand{\HH}{{\mathcal{H}}}

\newcommand{\tr}{\operatorname{tr}}

\newcommand{\pp}[1]{\begin{pmatrix} #1 \end{pmatrix}}

\newcommand{\bb}{\begin{eqnarray}}
\newcommand{\ee}{\end{eqnarray}}
\newcommand{\eee}{\nonumber\end{eqnarray}}

\newcommand{\rxy}[1]{{\begin{xy}0;<2mm,0mm>:<0mm,2mm>::0;0,#1\end{xy}}}

\newcommand{\minibox}[2]{{\fbox{\begin{minipage}{#1}\begin{center}#2
\end{center}\end{minipage}}}}

\newcommand{\Photo}{}

\begin{document}
\vspace*{4cm}
\title{NONCOMMUTATIVE GEOMETRY IN THE LHC-ERA}

\author{ C.A. STEPHAN }

\address{Institut f\"ur Mathematik, Universit\"at Potsdam, Am Neuen Palais 10,\\
14469 Potsdam, Germany}

\maketitle\abstracts{
Noncommutative geometry  allows to unify the basic building blocks of
particle physics, Yang-Mills-Higgs theory and General relativity, into a single geometrical framework. 
The resulting effective theory constrains the couplings of the particle model and reduces the number of 
degrees of freedom.
After briefly introducing the basic ideas of noncommutative geometry, I will present its predictions
for the Standard Model (SM) and the few known models beyond the Standard Model. Most of these models, including the Standard Model, 
are now ruled out  by LHC data. But interesting extensions of the Standard Model which agree with the presumed 
Standard Model Scalar (SMS) mass  and predict new particles are still very much alive and await further experimental data.}

\section{Introduction}
The aim of noncommutative geometry is to unify General Relativity and the Standard Model (or suitable extensions of the Standard Model) 
on the same geometrical footing. This means to describe gravity
and the electro-weak and strong forces as gravitational forces of a unified space-time. 
A first observation on  the structure of General Relativity shows that
gravity emerges as a pseudo-force associated to the space-time (=  a manifold $M$) symmetries, 
i.e. the diffeomorphisms on $M$. If one tries to put the Standard Model into the same scheme, one cannot 
find a manifold  within classical differential geometry  which could be the equivalent of space time.
A second observation is that one can find an equivalent description of space(-time)
when trading the differential geometric description of the Euclidean space(-time) manifold $M$ with metric
$g$ for the algebraic description of a spectral triple $(C^\infty(M), \slashed{\partial},\mathcal{H})$. 
A spectral triple~\cite{C94} consists of the following entities:
\begin{itemize}
\item[$\bullet$] an algebra $\mathcal{A}$ $(=C^\infty(M))$, the equivalent of the topological space $M$
\item[$\bullet$] a Dirac operator $\mathcal{D}$ $(=\slashed{\partial})$, the equivalent of the metric $g$
on $M$
\item[$\bullet$] a Hilbert space $\mathcal{H}$, on which the algebra is faithfully represented and on which the Dirac operator acts. It contains the fermions, i.e. in
the space(-time) case, it is the Hilbert space of Dirac 4-spinors.
\item[$\bullet$] a set of axioms~\cite{C96} to ensure a consistent description of
the geometry
\end{itemize}
This setup allows to describe General Relativity in terms of spectral triples. Space(-time) is replaced by the
algebra of $C^\infty$-functions over the manifold and the Dirac operator plays a double role: 
it is the algebraic equivalent of the metric and it gives the dynamics of the Fermions.
The Einstein-Hilbert action is replaced by the spectral action~\cite{CC97}, which is the simplest invariant 
action given by the number of eigenvalues of the Dirac operator up to a cut-off energy. 
For readers interested in a more thorough introduction to the field we recommend~\cite{Sc05a} and~\cite{DS12}.

\noindent
Connes' key observation is that the geometrical notions of the spectral triple remain
valid, even if the algebra is {\it noncommutative}. A natural way of achieving
noncommutativity is done by multiplying the function algebra $C^\infty (M)$  with  a
sum of real, complex or quaternionic matrix algebras
\begin{equation}
\mathcal{A}_f = M_1(\mathbb{K}) \oplus M_2(\mathbb{K}) \oplus \dots
\label{intalg}
\end{equation}
These algebras represent {\it internal spaces} which
have the $O(N)$-, $U(N)$- or $SU(N)$-type Lie groups as their symmetries
that one would like to obtain as gauge groups in particle physics. The choice
\begin{equation}
\mathcal{A}_{SM} = \mathbb{C} \oplus M_2(\mathbb{C}) \oplus M_3(\mathbb{C})
\oplus \mathbb{C}
\label{Standard Model}
\end{equation}
allows to construct the Standard Model and can be justified by different classification
approaches~\cite{ISS04,JS08,CC08}.
The combination of space(-time) and internal space
into a product space, together with the spectral action~\cite{CC97},
unifies General Relativity and the Standard Model as classical field theories:
\begin{center}
\begin{tabular}{c}
\\
\rxy{
,(-20,0)*{
\minibox{4cm}{Almost-Commutative \\ Spectral Triple \\ $\mathcal{A}= C^\infty(M) \otimes
\mathcal{A}_{SM}$}
}
,(20,0)*{
\minibox{4cm}{Int.+Ext. Symmetries \\ = Diff$(M) \ltimes$ \\ $U(1) \times SU(2) \times SU(3)$}
}
,(0,-20)*{
\minibox{5cm}{Spectral Action = \\ E-H Act.+Cosm.Const. \\ + Stand. Model Action }
}
,(-15,-6);(-1,-15)**\dir{-}?(0)*\dir2{<}
,(15,-6);(1,-15)**\dir{-}?(1)*\dir2{>}
,(-9,0);(9,0)**\dir{-}?(0)*\dir2{<}
,(0,3)*{\mbox{act on}}
,(18,-10)*{\mbox{leave invariant}}
,(-14,-10)*{\mbox{Dynamics}}
}
\\ \\
\end{tabular}
\end{center}
The  Standard Model in the noncommutative geometry setting automatically produces:
\begin{itemize}
\item[$\bullet$] The combined General Relativity and Standard (Particle) Model action
\item[$\bullet$] A cosmological constant
\item[$\bullet$] The SMS boson with the correct quartic SMS potential
\end{itemize}
The Dirac operator turns out to be one of the central objects and plays a
multiple role:

\begin{center}
\begin{tabular}{c}
\rxy{
,(0,1)*{\minibox{2.5cm}{Dirac Operator}}
,(-25,-15)*{\minibox{2.6cm}{SMS \& Gauge \\ Bosons}}
,(25,-15)*{\minibox{2.6cm}{Metric of $M$,\\ Internal Metric}}
,(0,-15)*{\minibox{3.5cm}{Particle Dynamics, \\ Ferm. Mass Matrix}}
,(-1,-3);(-25,-9)**\dir{-}?(1)*\dir2{>}
,(1,-3);(25,-9)**\dir{-}?(1)*\dir2{>}
,(0,-3);(0,-9)**\dir{-}?(1)*\dir2{>}
}
\\ \\
\end{tabular}
\end{center}
Up to now one could conclude that noncommutative geometry consists merely in a fancy, 
mathematically involved reformulation of the Standard Model.
But the choices of possible Yang-Mills-Higgs models that fit into
the noncommutative geometry framework are limited. Indeed the geometrical setup leads
already to a set of restrictions on the possible particle models:
\begin{itemize}
\item mathematical axioms  $=>$ restrictions on particle content
\item symmetries of finite space $=>$ determines gauge group
\item representation of matrix algebra  $=>$ representation of gauge group

(only fundamental and adjoint representations)
\item Dirac operator $=>$ allowed mass terms / scalar fields
\end{itemize}
Further constraints come from
the Spectral Action which results in an effective action valid at a
cut-off energy $\Lambda$~\cite{CC97}. This effective 
action comes with a set of constraints on the particle model
couplings, also valid at $\Lambda$, which reduce the 
number of free parameters in a significant way.

\noindent
The Spectral Action contains two parts. A fermionic part $( \Psi, \mathcal{D} \Psi )$
which is obtained by inserting the Dirac operator into the scalar product
of the Hilbert space and a bosonic part $S_\mathcal{D} ( \Lambda)$
which is just the number of eigenvalues of the Dirac operator up to
the cut-off:
\begin{itemize}
\item $( \Psi, \mathcal{D} \Psi )$ = fermionic action  includes Yukawa couplings 
 \& fermion--gauge boson interactions
\item $S_\mathcal{D} ( \Lambda)$ = the bosonic action given by the
 number of eigenvalues of $\mathcal{D}$ up to cut-off $\Lambda$

\hskip1.2cm = the Einstein-Hilbert action + a Cosmological  Constant 

\hskip1.5cm + the full bosonic particle model action + constraints at $\Lambda $

\end{itemize}
The bosonic action $S_\mathcal{D} ( \Lambda)$ can be calculated explicitly 
using the well known heat kernel expansion~\cite{CC97}. Note that $S_\mathcal{D} ( \Lambda)$ is manifestly gauge invariant and also invariant under
the diffeomorphisms of the underlying space-time manifold.

\noindent
For the Standard Model the internal space is taken to be the matrix algebra
 $\mathcal{A}_f=  \mathbb{C} \oplus M_2(\mathbb{C} ) \oplus M_3(\mathbb{C} )
\oplus \mathbb{C}$. The symmetry  group of this discrete space
is given by the group of (non-abelian) unitaries of $\mathcal{A}_f$: 
$U(2)  \times U(3)$. This leads, when properly lifted to the Hilbert space,
to the Standard Model gauge group  $U(1)_Y \times SU(2)_w  \times SU(3)_c$.
It is a remarkable fact that the Standard Model fits so well into the noncommutative geometry framework!

\noindent
Calculating from the geometrical data the Spectral Action leads to the
following boundary conditions on the Standard Model parameters at the cut-off $\Lambda$
\begin{equation}
\frac53 g_1^2 =\, g_2^2=\, g_3^2=\,\frac{Y_2^2}{G_2} \,\frac{\lambda}{24}\,= \,\frac{1}{4}\,Y_2 
\label{conSM}
\end{equation}
Where  $g_1$, $g_2$ and $g_3$ are the $U(1)_Y$, $SU(2)_w$ and $SU(3)_c$  gauge couplings,
$\lambda$ is the quartic scalar coupling,
$Y_2$ is the sum of all Yukawa couplings squared
and $G_2$ is the sum of all Yukawa couplings to the fourth power.

\noindent
Assuming the Big Desert and the stability of the theory under the
flow of the renormalisation group equations we can 
deduce that $g_2^2(\Lambda) =g_3^2 (\Lambda) $ at $\Lambda = 1.1 \times 10^{17}$ GeV 
Having thus fixed the cut-off scale $\Lambda$ we can
use the remaining constraints to determine the low-energy value
of the quartic SMS coupling and the top quark Yukawa coupling (assuming
that it dominates all Yukawa couplings). 
This leads to the following conclusions:
\begin{itemize}
\item $\tfrac53 \, g_1^2 \neq g_2^2$ at $\Lambda$
\item $m_{\rm SMS} = 168.3 \pm 2.5$ GeV
\item $m_{\rm top} < 190$ GeV
\item no $4^{th}$ Standard Model generation
\end{itemize}
So the boundary conditions \eqref{conSM} of the Standard Model cannot be fulfilled for the gauge couplings. 
This is of course a well known fact, since the conditions coincide with the $SU(5)$-grand unified case.
Note that the prediction for the SMS mass of $\sim 170$ GeV is also ruled out.
This value has recently been shown to be too high~\cite{ATLAS12,CMS12} since the 
SMS has a mass of $\sim 125$ GeV. Thus it seems
plausible to consider also models beyond the Standard Model.

\noindent
Although the Standard Model takes a prominent place~\cite{ISS04,JS08,CC08} 
within the possible models of almost-commutative geometries one can to go further and construct models beyond 
the Standard Model. The techniques from the
classification scheme developed in~\cite{ISS04} were used to enlarge the Standard Model~\cite{St06a,St07,SS07,St09}, but
most of these models~\cite{St06a,St07,SS07} suffer from a similar shortcoming as the Standard Mode: The mass of the SMS is in 
general too high compared to the experimental value.
Here the model in~\cite{St09} will be of central interest, since it predicted approximately the correct SMS mass.

\noindent
In the case of finite spectral triples of KO-dimension six~\cite{B07,C06}
a different classification leads to more general versions of the Standard Model algebra~\cite{CC08}, under some extra assumptions. 
Considering the first order axiom  as being dynamically imposed on the spectral triple one finds a Pati-Salam
type model~\cite{CCS13}. From the same geometrical basis one can promote the Majorana mass of the neutrinos to a scalar field~\cite{CC12,DLM13} which allows to lower the SMS mass to its experimental value.
\section{The model}
The model we are investigating  extends the Standard Model~\cite{CC97} by $N$ generations of  chiral $X_1$- and $X_2$-particles and 
vectorlike $V_{c/w}$-particles. It is a variation of the model in~\cite{St09} and the model in~\cite{St13} . For details of the
following calculations as well as the construction of the spectral triple we refer the reader to~\cite{St13}. In particular  its 
Krajewski diagram is depicted in figure 4~\cite{Str13}. The necessary computational adaptions 
to the model in this publication are straightforward.  The gauge group of the Standard Model is enlarged by
an extra $U(1)_X$ subgroup, so the total group is $G=U(1)_Y \times SU(2)_w \times SU(3)_c \times U(1)_X$. 
The Standard Model particles are neutral with respect to the $U(1)_X$ subgroup while the $X$-particles 
are neutral with respect to the  Standard Model subgroup $G_{SM}=U(1)_Y \times SU(2)_w \times SU(3)_c $. 
Furthermore the model contains two scalar fields: a scalar field in the SMS representation and a new scalar field 
carrying only a $U(1)_X$ charge. They induce a symmetry breaking mechanism $G\to U(1)_{em} \times SU(3)_c$.
\medskip

\noindent
The Hilbert spaces of the new fermions and the new scalar field expressed in terms of their representations are
\bb
&& \HH_{X_1,l}^p \oplus \HH_{X_2,l}^p  = \bigoplus_1^N [(0,1,1,+1) \oplus (0,1,1,0)], 
\nonumber \\
&& \HH_{X_1,r}^p \oplus \HH_{X_2,r}^p = \bigoplus_1^N  [(0,1,1,0) \oplus (0,1,1,+1)],
\nonumber \\
&& \HH_{V_c,l}^p \oplus \HH_{V_w,l}^p = \HH_{V_c,r}^p \oplus \HH_{V_w,r}^p  = \bigoplus_1^N [(-\tfrac13,1,\bar 3,\tfrac12) \oplus (0,\bar 2,1,\tfrac12)]
\label{repsfermions} \\
&& \HH_{\varphi} = (0,1,1,-1)
\ee
Here we chose the standard normalisations of~\cite{MV83,MV84,MV85} for the representation, i.e. the right-handed electron has hypercharge $Y_e=-1$.

\noindent
The Lagrangians for the $X$-particles, 
\bb
\mathcal{L}_{\nu X_1} &=& \left( (g_{\nu X_1}) \, \psi_{\nu,r}^* \, \gamma_5 \, \varphi \, \psi_{X_1,l} \, +  \, {\rm h.c.} \right)
\nonumber \\
\mathcal{L}_{X_i} &=& ( \psi_{X_i,l}^* , \psi_{X_i,r}^*) D_{X_i}  \pp{\psi_{X_i,l} \\ \psi_{X_i,r} }
\; + \;  \left(  \, \psi_{X_i,l}^* \, \gamma_5 \, g_{X_i} \psi_{X_i,r} \, +  \, {\rm h.c.} \right)
\nonumber \\
\mathcal{L}_{X_2,\Gamma} &=& \psi_{X_2,r}^* \, \gamma_5 \, \Gamma_{X_2} \,  \psi_{X_2,r} 
\label{XLagrangian}
\ee
contain the term $\mathcal{L}_{\nu X_1}$ coupling $X_1$ to the right-handed neutrinos, the ordinary Dirac and Yukawa term
$\mathcal{L}_{X_i}$ for both $X_{1/2}$-particle species and a Majorana mass term $\mathcal{L}_{X_2,\Gamma}$ for the right-handed
$X_2$-particles. The fermionic Lagrangian of the $V_c$- and the $V_w$- particles is
\bb
\mathcal{L}_{V_{c/w}} &=&  ( \psi_{V_{c/w},l}^* , \psi_{V_{c/w},r}^*) D_{V_{c/w}}  \pp{\psi_{V_{c/w},l} \\ \psi_{V_{c/w},r} }
\; + \; \left(  \psi_{V_{c/w},l}^* \, \gamma_5 \, \M_{V_{c/w}} \psi_{V_{c/w},r} \, +  \, {\rm h.c.} \right).
\label{VLagrangian}
\ee
Together with usual fermionic Lagrangian of the Standard Model, $\mathcal{L}_{SM,f}$ they give the fermionic part of the theory. 
The Dirac operators of the form  $D_X$ and $D_V$ in \eqref{XLagrangian} and \eqref{VLagrangian} are the  Dirac operators
with the respective gauge covariant derivatives.

\noindent
The Yang-Mills Lagrangian for the  Standard Model subgroup $G_{SM}$ takes again its usual form  and for $U(1)_X$ we have the standard Yang-Mills-Lagrangian
\bb
\mathcal{L}_{U(1)_X} = \frac{1}{4 g_4^2}  \sum_{i,j=1}^4 (B^X)_{ij} (B^X)_{ij} 
\ee
where $B^X$ is the field strength tensor for the $U(1)_X$-covariant derivative. The Lagrangian of the scalar fields  
\bb
\mathcal{L}_{H,\varphi} =  |\nabla^{H} H |^2  +  |\nabla^{\varphi} \varphi|^2 - \mu_H^2 |H|^2 - \mu_\varphi^2 |\varphi|^2 + 
\frac{\lambda_1}{6} |H|^4 + \frac{\lambda_2}{6} |\varphi|^4 + \frac{\lambda_3}{3} |H|^2 \, |\varphi|^2 
\label{ScalarLagrangian}
\ee
contains an interaction term of the $H$-field and the $\varphi$-field.
The normalisation is again chosen according to~\cite{MV83,MV84,MV85} in order to obtain Lagrangians of the form
$(1/2) (\partial \varphi)^2 -(\mu^2/2) \varphi^2 + (\lambda/24) \varphi^4$ for the real valued fields.
Putting everything together the Dirac inner product  $\langle \psi, D_\Phi \, \psi \rangle$  and the Spectral Action
provide the Lagrangian
\bb
\mathcal{L}_{full} = \mathcal{L}_{SM,f} + \mathcal{L}_{SM,YM} + \mathcal{L}_{U(1)_X} + \mathcal{L}_{H,\varphi} + \mathcal{L}_{\nu X_1} +\mathcal{L}_{X_1}
+\mathcal{L}_{X_2} + \mathcal{L}_{X_2,\Gamma} + \mathcal{L}_{V_c} + \mathcal{L}_{V_w}.
\label{totalLagrangian}
\ee
The Spectral Action  does not only supply the  bosonic part of the Lagrangian \eqref{totalLagrangian} but also
relations among the free parameters of the model. These relations serve as boundary conditions for the renormalisation group flow
at the cut-off scale $\Lambda$.  The relation for the gauge couplings $g_1$, $g_2$,
$g_3$ and $g_4$ at the cut-off scale $\Lambda$ are:
\bb
\boxed{
\sqrt{\frac{80 + \tfrac43 N}{48 + 8 N}} \; g_1(\Lambda) = g_2(\Lambda)=g_3(\Lambda) = \sqrt{\frac{36 N}{48 + 8 N}} \; g_4(\Lambda) 
}
\label{AllUatLambda}
\ee
We notice the deviation of these relations compared to the case of the Standard Model \eqref{conSM}~\cite{CC97}. 
For the Yukawa couplings the Spectral Action  implies the boundary conditions~\cite{T03}
\bb
\boxed{
Y_2(\Lambda) = Y_X(\Lambda)= (4+ \tfrac23 N) \, g_2(\Lambda)^2 
}
\label{YukawasatLambda}
\ee
where we have introduced the squared traces
\bb
Y_2 &:=& 3 \tr(g_u^* g_u) + 3 \tr(g_d^* g_d) + \tr(g_e^* g_e)  + \tr(g_{\nu}^* g_{\nu})
\nonumber \\
Y_X&:=& \tr( g_{\nu X_1}^* g_{\nu X_1}) + \tr( g_{X_1}^* g_{X_1}) +\tr( g_{X_2}^* g_{X_2}). 
\label{squaredsums} 
\ee
in terms of the Yukawa coupling matrices. With the traces of Yukawa matrices to the fourth power,
\bb
G_2 &:=& 3 \tr[(g_u^* g_u)^2] + 3 \tr[(g_d^* g_d)^2] + \tr[(g_e^* g_e)^2]  + \tr[(g_{\nu}^* g_{\nu})^2]
\nonumber \\
G_X&:=&  \tr[(g_{\nu X_1}^* g_{\nu X_1})^2]+\tr[( g_{X_2}^* g_{X_2})^2]  +\tr[( g_{X_3}^* g_{X_3})^2] 
+2 \tr  (g_{X_1}^* g_{X_1} g_{\nu X_1}^* g_{\nu X_1})  
\nonumber \\
G_{\nu X_1} &:=& \tr (g_\nu^* g_\nu g_{\nu X_1}^* g_{\nu X_1}). 
\label{fourthpowersums}
\ee
the boundary conditions of  the quartic couplings read 
\bb
\boxed{
\lambda_1(\Lambda) =  g_2(\Lambda)^2 (24 + 4 N) \frac{ G_2(\Lambda)}{Y_2(\Lambda)^2}, \quad 
\lambda_2(\Lambda)= g_2(\Lambda)^2 (24 + 4 N) \frac{ G_X(\Lambda)}{Y_X(\Lambda)^2}
}
\label{lambda12atLambda}
\ee
and
\bb
\boxed{
\lambda_3(\Lambda) = g_2(\Lambda)^2 (24 + 4 N) 
\frac{ G_{\nu X_1}(\Lambda)}{Y_2(\Lambda)Y_X(\Lambda)} 
}
\label{lambda3atLambda}
\ee

\section{A numerical example \label{NE}}

We assume that we have $N=3$ generations of new particles since there are three known Standard Model generations.
This fixes the conditions of the gauge couplings at the cut-off scale $\Lambda$ in \eqref{AllUatLambda}. 
We also assume that the Standard Model Yukawa couplings are dominated by the top-quark coupling $y_t$ and the 
$\tau$-neutrino coupling $y_{\nu_\tau}$.
Assuming that the Yukawa couplings that involve the $X$-particles are dominated by the coupling of the $\tau$-neutrino to one generation of the 
$X_1$-particles $y_{\nu_\tau X_1^1}$ we get  for \eqref{squaredsums} and
\eqref{fourthpowersums}
\bb
Y_2 &\approx& 3 \, y_t ^2+ y_{\nu_\tau}^2, \qquad Y_X \approx y_{\nu_\tau X_1^1}^2, 
\nonumber \\
G_2 &\approx& 3 \, y_t ^4+ y_{\nu_\tau}^4, \qquad G_X \approx y_{\nu_\tau X_1^1}^4,  \qquad G_{\nu X_1} \approx y_{\nu_\tau}^2 y_{\nu_\tau X_1^1}^2. 
\eee
The Majorana masses of the neutrinos~\cite{JKSS07} and the Dirac masses of the $X_1$-particles we put to $m_{X_1} \sim 10^{14}$GeV. 
The Majorana masses of the $X_2$-particles play no essential r\^ole for this numerical analysis. It triggers a seesaw mechanism for the $X_2$-particles
but has no effect on the masses of the scalar fields since we neglect the Yukawa matrix $g_{X_2}$.
For the masses of the three $V_c$-particles we choose $m_{V_c} \sim 5.5 \times 10^{15}$GeV. This particular value of the $V_c$-particle masses 
allows to fulfill the boundary conditions \eqref{AllUatLambda}.
 
\noindent
Defining $x := g_{\nu_\tau}/g_t$ the boundary relations \eqref{AllUatLambda},  \eqref{YukawasatLambda},
\eqref{lambda12atLambda} and \eqref{lambda3atLambda} among the dimensionless parameters 
at the cut-off scale $\Lambda$ simplify to
\bb
\frac76 \; g_1^2 = g_2^2=g_3^2 = \frac32 \; g_4^2 =  \frac16 \, g_t^2 (3 + x^2)   = \frac16 \, y_{\nu_\tau X_1^1}^2
= \frac{1}{36} \frac{(3+ x^2)^2}{3+x^4} \, \lambda_1
= \frac{1}{36}\, \lambda_2
= \frac{1}{36} \frac{3+x^2}{x^2} \, \lambda_3
\label{simpleboundary}
\ee
So the remaining free parameter for this particular point in parameter space, the ratio $x$,  has to be chosen such that the low energy value of $g_t$ 
coincides with its experimental value given by the experimental value of the top-quark mass. 
All normalisations are chosen as in~\cite{JKSS07}, i.e. the Standard Model fermion masses are given by $m_f = \sqrt{2} (g_f/g_2) m_{W^\pm}$.

\begin{figure}
\includegraphics[scale=0.4]{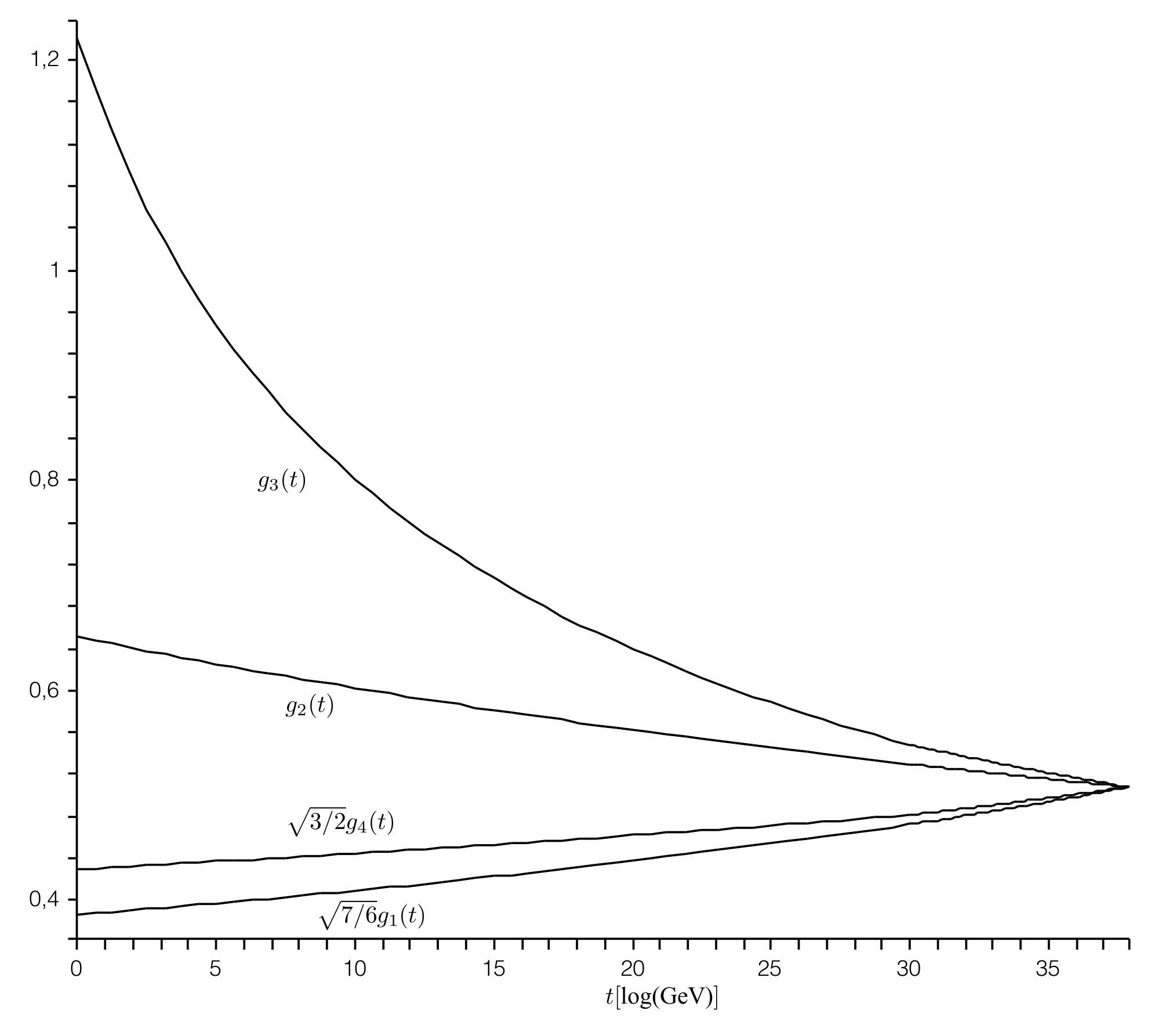}
\includegraphics[scale=0.4]{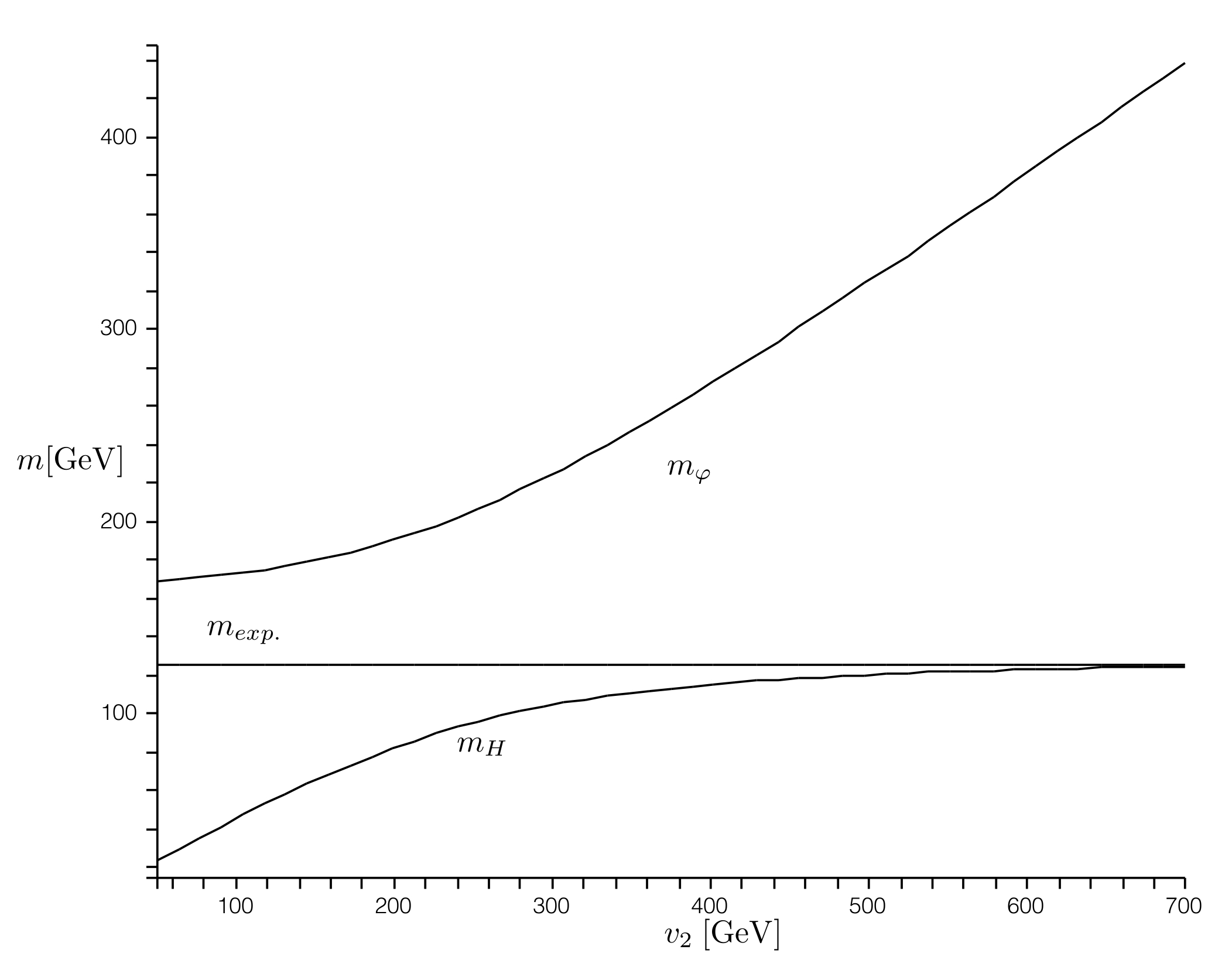}
\caption{Left:Running of the gauge couplings with normalisation factors according to the high-energy boundary conditions \eqref{simpleboundary}.
Right: Example for the mass eigenvalues $m_H$ and $m_\varphi$  of the scalar fields $H$ and $\varphi$ with respect
to $v_2$. Here the top quark mass is taken to be $m_t (m_Z) =173.5$ GeV and the experimental SMS mass $m_{exp.}
= 125.5$ GeV. We obtain a heavy scalar with $m_\varphi = 320$ GeV and a $U(1)_X$-scalar boson mass 
$m_{Z_X} = 172$ GeV. }
\label{massEWs1}
\end{figure}

\noindent
Let us determine the cut-off scale $\Lambda$ one-loop renormalisation group equations~\cite{MV83,MV84,MV85} for $g_1$, $g_2$ and $g_3$. 
As experimental low-energy values we take~\cite{PDG12}
\bb
g_1(m_Z)=0.3575, \quad g_2(m_Z)= 0.6514, \quad g_3(m_Z)= 1.221. 
\eee
The one-loop renormalisation group analysis shows that the high-energy boundary conditions for the gauge couplings in  \eqref{simpleboundary}
\bb
\frac76 \; g_1(\Lambda)^2 = g_2(\Lambda)^2=g_3(\Lambda)^2
\eee
can be met for $\Lambda \approx 2.6 \times 10^{18}$ GeV.  The running of the gauge couplings is shown in figure \ref{massEWs1}. 

\noindent
Both scalar fields  have nonzero vacuum expectation values, $|\langle H \rangle| = v_H/\sqrt{2} \neq  0$ and
$|\langle \varphi \rangle| = v_\varphi/\sqrt{2} \neq 0$. 
With  the $W^\pm$ boson mass,  $m_{W^{\pm}}=(g_2/2) \, v_H$, we can determine the  vacuum expectation value $v_H$ of
the first scalar field and obtain with $m_{W^{\pm}}=80.34$ GeV~\cite{PDG12} its experimental value $v_H = 246.8$ GeV.
In the numerical analysis it turns out that the errors are dominated by the experimental uncertainties of the top quark mass and
the SMS mass. Therefore we neglect the experimental uncertainties of the $W^\pm$ bosons and of the gauge couplings
$g_1$, $g_2$ and $g_3$.
The vacuum expectation value of $\varphi$ is a free parameter and is
determined by $\mu_\varphi$. The mass matrix is not diagonal in the weak basis, but the
mass eigenvalues can be calculated to be~\cite{EK07}
\bb
m_{\phi_H,\phi_\varphi} = \frac{\lambda_1}{6} \, v_H^2 + \frac{\lambda_2}{6} \, v_\varphi^2
\pm \sqrt{ \left(\frac{\lambda_1}{6} \, v_H^2 - \frac{\lambda_2}{6} \, v_\varphi^2 \right)^2
+ \frac{\lambda_3^2}{9} \, v_H^2 v_\varphi^2}.
\label{masseigenvals}
\ee
To obtain the low energy values of the quartic couplings $\lambda_1$, $\lambda_2$ and $\lambda_3$ we evolve their values at the cut-off $\Lambda$ fixed by the boundary conditions \eqref{simpleboundary}.
The top-quark mass has the experimental value $m_t =173.5 \pm 1.4$ GeV~\cite{PDG12} which determines the paramter $x=2.145 \pm 0.065$.
As a very recent input we have the SMS mass $m_{\Phi_H}=125.6 \pm 1.2$ GeV with the maximal combined experimental uncertainties of ATLAS~\cite{ATLAS12} and 
CMS~\cite{CMS12}. We identify the SMS mass with the lower mass eigenvalue in \eqref{masseigenvals}. As it turns out, this is a reasonable 
assumption as long as the vacuum expectation value $v_\varphi$ is larger then the $v_H$.

\noindent
With the above values of the top quark and the $W^\pm$ boson masses and the previous results on the gauge couplings we obtain from 
the running of the couplings, form the whole set of boundary conditions \eqref{simpleboundary} at $\Lambda$  
and from the mass eigenvalue equation
\eqref{masseigenvals} that $v_\varphi = 595 \pm 159$ GeV. The error in the vacuum expectation value is rather large due to the flat slope of the curve of the
light eigenvalue, see figure \ref{massEWs1}. The flat slope amplifies the experimental uncertainties of the top quark mass and
the SMS mass. We notice that $v_\varphi$ is larger then $v_H$ but of the same order of magnitude. This 
justifies the assumption to run all three quartic couplings down to low energies and also ensures that the vacuum is stable up to the cut-off scale
$\Lambda \approx 2.6 \times 10^{18}$  GeV, see~\cite{MEGLS12} and references therein.
For the second mass eigenvalue we find $m_{\Phi_\varphi} = 381 \pm 90$ GeV. For the mass of the new gauge boson $Z_X$ associated to the
broken $U(1)_X$ subgroup we get $m_{Z_X} =  209 \pm 56$ GeV where we took the value of $g_4(250 \; {\rm GeV})\approx 0.351$ as a very good
approximation and neglected further uncertainties of $g_4$. Note that the masses of the gauge and the scalar sector are lower than in the 
related model~\cite{St13}.

\noindent
To get a rough estimate of the masses of the $X_1$- and $X_2$-particles we note that the $g_{\nu_\tau X_1^1} \sim 1$. So if we assume that
$g_{\nu_\tau X_1^1}$ dominates the sum of the squared Yukawa coupling $Y_X$ with an accuracy of a few percent, it is reasonable to take  $g_{X_{1/2}^i} \leq 0.1$ for the remaining Yukawa
couplings. This leads to $X_{1}$-particle masses of the order 50 GeV, or less. If we assume that the Majorana masses of the $X_2$-particles 
$(\Gamma_{X_2})_{ij} \gg  v_\varphi$ then we expect three mass eigenstates of order $|(\Gamma_{X_2})|$ and three of order 
$v_\varphi^2/|(\Gamma_{X_2})|$. These particles could therefore populate lower mass scales.

\noindent
Let us end this short note with some open questions. We have only performed a one-loop renormalisation group analysis. Yet
the two-loop effects on the running of the top-quark mass are of the order of $15 \%$. In view of the sensibility of the 
scalar masses to the top-quark mass, how stable are the numerical results with respect to two-loop effects?
Furthermore how stable are the results with respect to variations in the parameter space?
From the experimental point of view  the question is of course the detectability of the new particles (at the LHC?) and whether the dark sector 
contains suitable dark matter candidates. One would expect more kinetic mixing  of $U(1)_X$ to the hypercharge 
group $U(1)_Y$ compared to the model~\cite{St13} due to the fact that here the $V_c$-particles may have a different masses and 
therefore mix the abelian groups.
This poses the question how $Z'$-like the $Z_X$-boson are and whether they are already excluded (at least for the
mass range explored in this publication). The model certainly has a rich phenomenology and is one of the few known models
to be consistent with (most) experiments, with the axioms of noncommutative geometry and with the boundary conditions imposed by 
the Spectral Action.

\section*{Acknowledgments}

The author appreciates financial support from the SFB 647: {\it Raum-Zeit-Materie}
funded by the Deutsche Forschungsgemeinschaft.

\end{document}